\documentclass[twocolumn]{aastex62}

\hypersetup{linkcolor=red}
\graphicspath{{./}{Figures/}}

\newcommand\st{\bgroup\markoverwith
{\textcolor{magenta}{\rule[0.5ex]{2pt}{1pt}}}\ULon}

\usepackage{amsmath, bm, hyperref, ulem}
\usepackage{xcolor, colortbl, array, makecell} 
\newcolumntype{?}[1]{!{\vrule width #1}}
\usepackage{savesym} \savesymbol{tablenum}
\usepackage[detect-weight=true, detect-family=true]{siunitx}
\usepackage{hyperref}
\restoresymbol{SIX}{tablenum}
\usepackage[T1]{fontenc}
\usepackage[overload]{textcase}
\usepackage{xcolor}
\usepackage{totcount}

\newtotcounter{citnum} 
\def\oldbibitem{} \let\oldbibitem=\bibitem
\def\bibitem{\stepcounter{citnum}\oldbibitem}
\graphicspath{{./}{Figures/}}
\shorttitle{Twin Planets} \shortauthors{McGruder et al.}

\begin{document}

\title{The {\it Similar Seven}: 
A set of very-alike exoplanets to test correlations between system parameters and atmospheric properties }

\correspondingauthor{Chima D. McGruder} \email{chima.mcgruder@cfa.harvard.edu}

\author[0000-0002-6167-3159]{Chima D. McGruder}\altaffiliation{NSF Graduate Research Fellow}\affiliation{Center for Astrophysics ${\rm \mid}$ Harvard {\rm \&} Smithsonian, 60 Garden St, Cambridge, MA 02138, USA}

\author[0000-0003-3204-8183]{Mercedes L\'opez-Morales} \affiliation{Center for Astrophysics ${\rm \mid}$ Harvard {\rm \&} Smithsonian, 60 Garden St, Cambridge, MA 02138, USA}

\author[0000-0002-9158-7315]{Rafael Brahm}
\affiliation{Facultad de Ingenier\'ia y Ciencias, Universidad Adolfo Ib\'a\~nez, Av.\ Diagonal las Torres 2640,
Pe\~nalol\'en, Santiago, Chile}
\affiliation{Millennium Institute for Astrophysics, Santiago, Chile}
\affiliation{Data Observatory Foundation}

\author[0000-0002-5389-3944]{Andr\'es Jord\'an}\affiliation{Facultad de Ingenier\'ia y Ciencias, Universidad Adolfo Ib\'a\~nez, Av.\ Diagonal las Torres 2640, Pe\~nalol\'en, Santiago, Chile}\affiliation{Millennium Institute for Astrophysics, Santiago, Chile}
\affiliation{Data Observatory Foundation}



\turnoffedit To turn off the edits, from https://journals.aas.org/revision-history/
\begin{abstract} 
Studies of exoplanetary atmospheres have found no definite correlations between observed high altitude aerosols and other system parameters. This could be, in part, because of the lack of homogeneous exoplanet samples for which specific parameters can be isolated and inspected. Here we present a set of seven exoplanets with very similar system parameters. We analyze existing photometric timeseries, Gaia parallax, and high-resolution spectroscopic data to produce a new set of homogeneous stellar, planetary, and orbital parameters for these systems. With this we confirm that most measured parameters for all systems are very similar, except for the host stars' metallicities and possibly high energy irradiation levels, which require UV and X-ray observations to constrain. From the sample, WASP-6b, WASP-96b and WASP-110b, have observed transmission spectra that we use to estimate their aerosol coverage levels using the Na I doublet 5892.9{\AA}. We find a tentative correlation between the metallicity of the host stars and the planetary aerosol levels. The trend we find with stellar metallicity can be tested by observing transmission spectra of the remaining planets in the sample. Based on our prediction, WASP-25b and WASP-55b should have higher levels of aerosols than WASP-124b and HATS-29b. Finally, we highlight how targeted surveys of alike planets like the ones presented here might prove key for identifying driving factors for atmospheric properties of exoplanets in the future and could be used as a sample selection criterium for future observations with e.g. JWST, ARIEL, and the next generation of ground-based telescopes.
\end{abstract}

\keywords{planets and satellites: atmospheres --- 
stars: activity; starspots --- techniques: spectroscopic; WASP-6b; WASP-25b; WASP-55b; WASP-95b; WASP-110b; WASP-124b, HATS-29b}


\section{Introduction} \label{sec:intro}
The upper atmospheres of about 100 exoplanets have been probed via transmission spectroscopy to date with HST, Spitzer, ground-based telescopes \edit2{\citep{NASA_archive_TranSpecs} \footnote{Accessed on 2022-05-29}}, and now JWST \citep{WASP39JWST2022}. Most observations suggest that atmospheric features are heavily muted by aerosols \citep[e.g.][]{Wakeford:2019}, with only a few planets showing little to no aerosol coverage \citep[e.g.][]{kirk:2019,Alam:2021,Ahrer:2022,McGruder2022}. 

The formation of aerosols in planetary atmospheres occurs via complex chemical and physical processes, which are not yet fully understood, but work is in progress \citep[see e.g.][]{Helling:2008,Marley:2013, Helling:2019,Gao2021}. For example, predictions link cloud formation in exoplanets to the chemical composition of their atmospheres, where \textit{seed} particles need to be lofted to high altitudes so gases can condense on them and form clouds \citep[e.g.][and references therein]{Helling:2008}. The composition, availability, and altitude of potential \textit{seed} particles has many dependencies, including the composition of the protoplanetary disk \citep[e.g.][]{Mordasini:2010}, and atmospheric differentiation, where heavier elements are expected to sink into the lower layers of the atmosphere \citep[e.g.][]{Helling:2019}. In the case of hazes, model predictions show that high altitude hazes can form in exoplanets via UV-driven photolysis \citep[e.g.][]{moses:2011, moses:2013}, and laboratory experiments have shown that UV radiation can form photochemical hazes in $H_2$ dominated atmospheres at temperatures between 600K--1500K \citep{Fleury:2019}. 

Understanding what system parameters drive the presence or absence of aerosols in exoplanetary atmospheres will be key to understanding their formation and evolution mechanisms. However, comparative studies of exoplanet atmospheres so far have yielded either no correlations between the atmospheric properties of the planets studied and other system parameters \citep[see e.g.][]{Sing:2016}, or only tentative correlations between planetary equilibrium temperature and atmospheric aerosol levels \citep{heng:2016,Stevenson:2016,fu:2017,Tsiaras:2018,Dymont2021}. Yet, degeneracies between system parameters and disagreeing observations leave any correlation uncertain \citep{Alam2020}.

Possible reasons for why correlations between observed exoplanet atmospheric properties and other system parameters have not yet been found are the fact that the parameter space being considered is too broad, with multiple independent parameters typically being fitted simultaneously (e.g. planetary equilibrium temperature, density, \edit1{log} surface gravity, planetary/host star metallicity, etc), and over wide range of values; as well as the parameter space being scarcely sampled, with typically one-to-no pairs of similar planets being examined. 

We present a way to potentially alleviate this problem by identifying groups of planets with similar properties, so that a reduced number of parameters can be isolated and compared against observed atmospheric features of the planets in detail. In particular, we have identified a group of seven gas giants with very similar parameters, except for the metallicity of their host stars and possibly their high energy irradiation levels. We use this group of planets to test whether aerosol properties are related to those parameters. 

Section \ref{sec:sample} describes the identification of the sample. Sections \ref{sec:DirParms} and \ref{sec:TransSpec} present our reanalysis of system parameters and observed transmission spectra for comparative purposes. Section \ref{sec:trends} presents the analysis of correlations between exoplanet aerosol level proxies and the host star metallicity and high energy irradiation levels. Finally, Section \ref{sec:Sum+Conc} summarizes our results.

\section{The Sample}\label{sec:sample}

Using the NASA Exoplanet Archive\edit2{\citep{PSCompPars}\footnote{Accessed on 2021-11-17}}, we compiled system parameters for all known exoplanets with an observed optical or near-infrared transmission spectrum. We compiled the planets mass, radius, orbital period and semi-major axis separation, as well as the mass, radius, effective temperature, and metallicity of the host stars. With those values in hand, we computed the planets gravity, density, equilibrium temperature, and their stellar insolation levels in earth units. 
We then searched in the compiled list for planets with very similar system parameters and compared their observed transmission spectra. 

\begin{figure*}[htb]
    \centering
    \includegraphics[width=.9\textwidth]{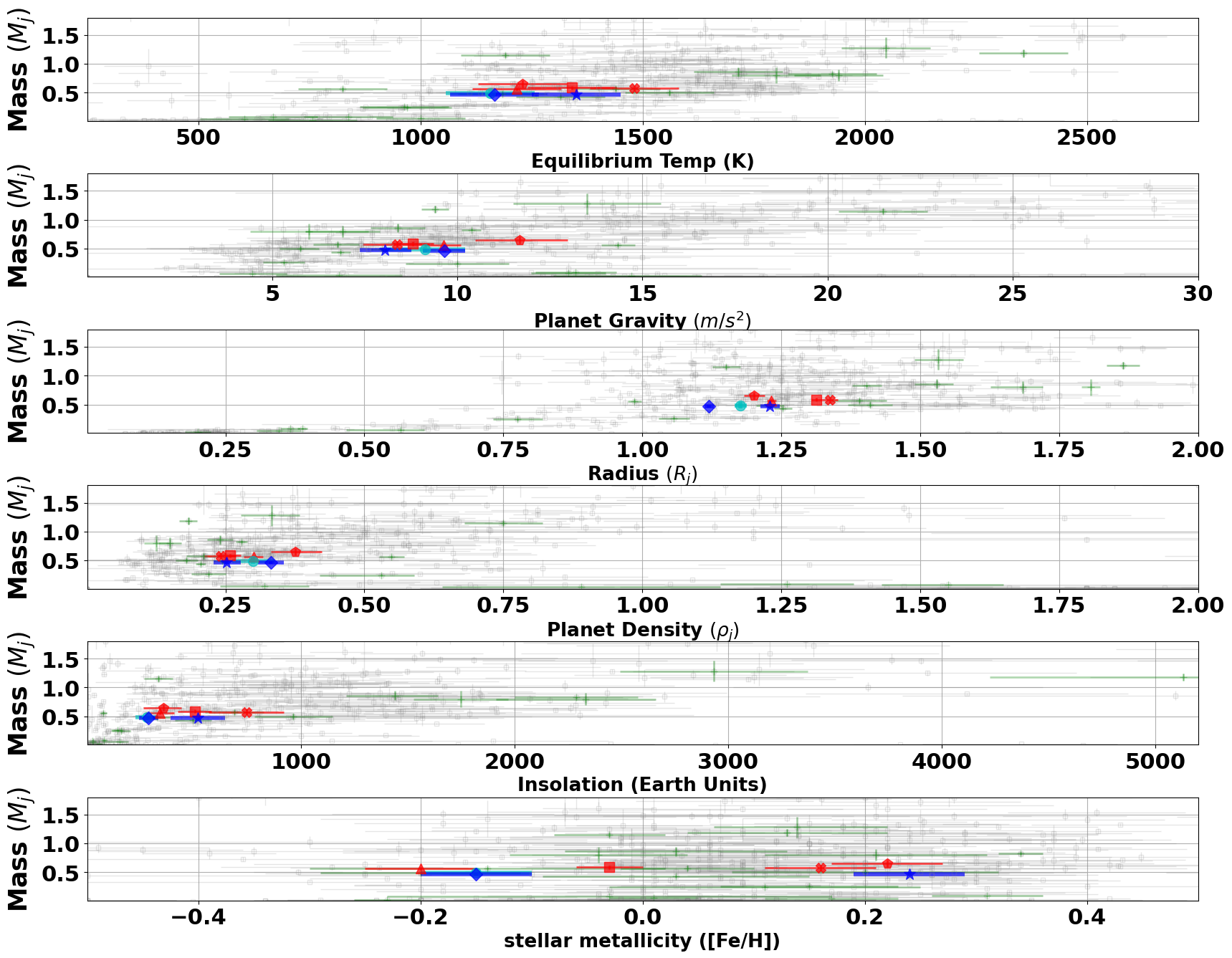}
    \includegraphics[width=.9\textwidth]{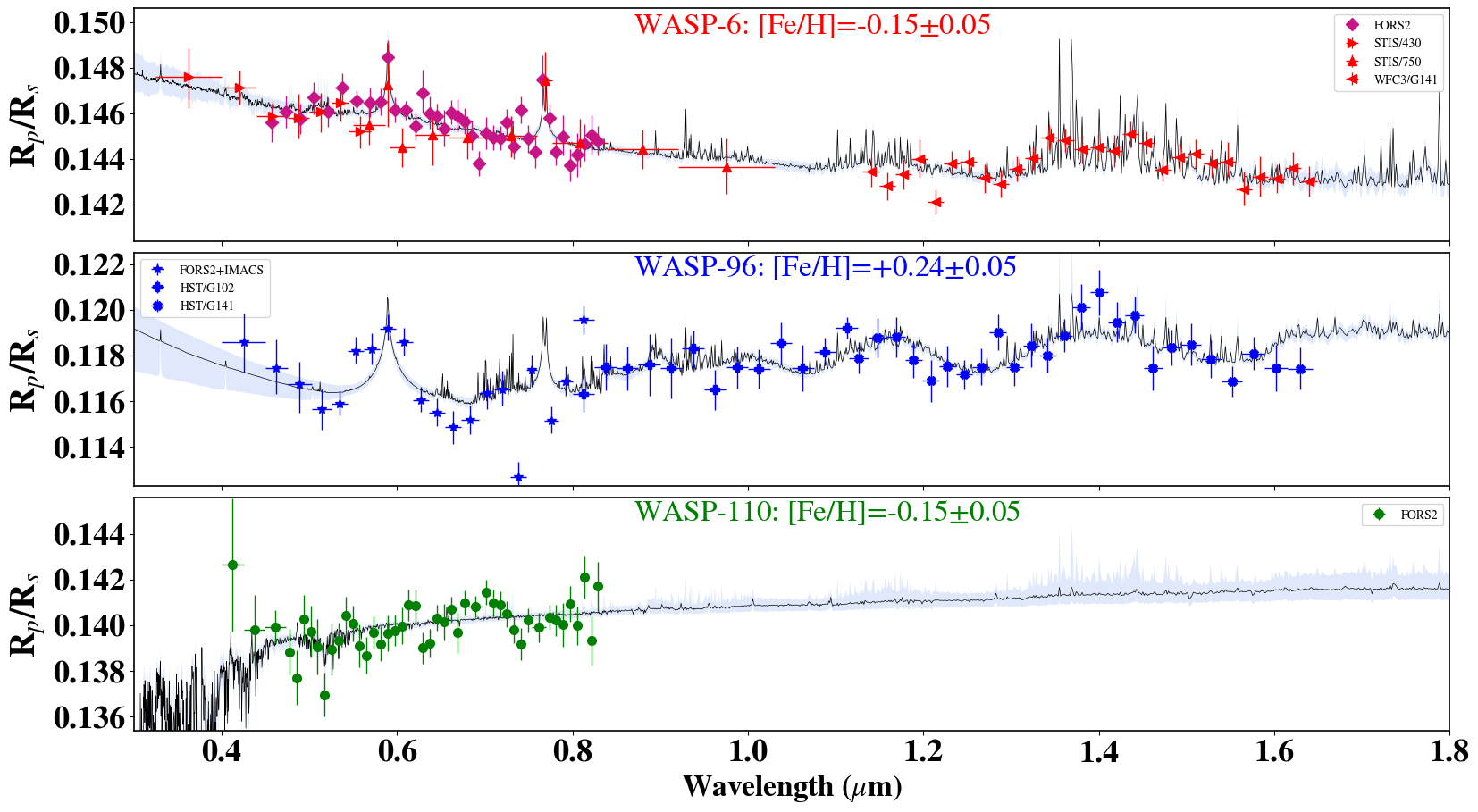}
    \caption{\textbf{Top:} Mass versus equilibrium temperature, gravity, radius, density, stellar irradiation, and star’s metallicity for WASP-6b (blue diamonds), WASP-96b (blue stars), WASP-110b (cyan circles), WASP-124b (red 'X'), WASP-55b (red squares), WASP-25b (red triangle), HATS-29b (red pentagon), the PanCET planets (green crosses), and all exoplanets with measured parameters (grey circles). \edit1{The Equilibrium Temperature, Insolation, and metallicity of WASP-110b and WASP-6b overlap (see Table \ref{tab:Sim7Params}), making it hard to see the cyan circles of WASP-110b underneath the blue diamonds of WASP-6b.} \textbf{Bottom:} Transmission spectrum of WASP-6b
    \citep[red/magenta,][]{Nikolov:2015, Carter2020}, WASP-96b \citep[blue,][]{Nikolov:2018, Yip:2021, Nikolov2022, McGruder2022}, and WASP-110b  \citep[green circles,][]{Nikolov2021}, with the \texttt{Platon} best-fit retrieval model (black line) and 1-$\sigma$ confidence interval (cyan shaded regions) overplotted. In this figure, the parameters for the \textit{similar seven} planets are the re-derived parameters discussed in Section \ref{sec:DirParms}.}
    \label{fig:W96vsW6_PlanPars} 
\end{figure*}

This is how we identified WASP-6b and WASP-96b, two hot jupiters with very similar masses, radii, $T_{eq}$, stellar parameters, and insolation levels (see top six panels of Figure \ref{fig:W96vsW6_PlanPars} and Table \ref{tab:Sim7Params}), but strikingly different transmission spectra (See bottom three panels of Figure \ref{fig:W96vsW6_PlanPars} and Section \ref{sec:TransSpec}). From the parameters available for each system, they only appear to differ significantly in the host star metallicity, reported in the NExSci database as [Fe/H] = -0.20$\pm$0.09 for WASP-6b \citep{Gillon2009}, and [Fe/H] = 0.14$\pm$0.19 for WASP-96b \citep{Hellier:2014}.

Next we compared the parameters of all known exoplanets, including those without atmospheric observations, to those of WASP-6b and WASP-96b, assuming the planets to be similar if all their planetary parameters listed above agreed within about 1$\sigma$. 
Using this strategy we found another five planets \textit{similar} to WASP-6b and WASP-96b: WASP-25b \citep{Enoch2011}, WASP-55b \citep{Hellier2012},  WASP-110b \citep{Anderson2014}, WASP-124b \citep{Maxted2016}, and HATS-29b \citep{Espinoza2016}. 
This is how we arrived to the \textit{similar seven} planets sample described in the remaining of the paper. The parameters of each system, re-derived homogeneously as described in the following section, are summarized in Table \ref{tab:Sim7Params}.

\section{Derivation of Homogeneous Star and Planet Parameters}\label{sec:DirParms}

The parameters used in Section \ref{sec:sample} to identify the sample were obtained from various literature sources. Therefore, to minimize potential biases and systematics between separate analyses, we re-derived the parameters of each system homogeneously. Table \ref{tab:Sim7Params} provides all the newly derived parameters, with the derivation processes of each set of parameters described below.

\newcommand\ChangeRT[1]{\noalign{\hrule height #1}}
\begin{deluxetable*}{|c|c|c|c|c|c|c|c|}[htb]
\caption{Stellar and planetary parameters of all Similar Seven Systems} 
    \label{tab:Sim7Params}
    \tablehead{\colhead{\bf{Param.}} & \colhead{\bf{WASP-6}} & \colhead{\bf{WASP-25}} & \colhead{\bf{WASP-55}} & \colhead{\bf{WASP-96}} & \colhead{\bf{WASP-110}} & \colhead{\bf{WASP-124}} & \colhead{\bf{HATS-29}}}
    \startdata 
    M$_*$ & 0.854$^{+0.027}_{-0.023}$ & 0.962$^{\pm0.027}_{\pm0.021}$ & 1.071$\pm0.025$ & 1.03$^{+0.031}_{-0.036}$ & 0.814$^{+0.014}_{-0.022}$ & 1.157$^{+0.016}_{-0.015}$ & 1.055$^{+0.036}_{-0.038}$ \\ \hline   
    R$_*$ & 0.79$^{+0.008}_{-0.009}$ & 0.884$^{+0.008}_{-0.009}$ & 1.09$^{+0.012}_{-0.013}$ & 1.055$^{+0.018}_{-0.017}$ & 0.853$\pm$0.011 & 1.074$^{+0.013}_{-0.015}$ & 1.066$^{+0.018}_{-0.019}$ \\ \hline
    T$_{eff}$ & 5438$\pm$50 & 5697$\pm80$ & 6096$\pm71$ & 5678$\pm$80 & 5392$\pm$50 & 6258$\pm$100 & 5769$\pm80$ \\ \hline
    log$_{10}(G_{*})$ & 4.565$^{+0.022}_{-0.016}$ & 4.529$^{\pm0.018}_{\pm0.014}$ & 4.393$\pm0.015$ & 4.404$^{+0.023}_{-0.026}$ & 4.487$^{+0.013}_{-0.018}$ & 4.439$^{+0.01}_{-0.009}$ & 4.406$\pm0.024$ \\ \hline
    $\rho_{*}$ & 1.73$^{+0.08}_{-0.07}$ & 1.391$^{+0.058}_{-0.048}$ & 0.822$^{+0.036}_{-0.035}$ & 0.876$^{+0.05}_{-0.054}$ & 1.31$^{+0.055}_{-0.062}$ & 0.933$^{+0.041}_{-0.036}$ & 0.869$^{+0.055}_{-0.054}$ \\ \hline
    [Fe/H] & -0.15$\pm$0.05 & -0.2$\pm0.05$ & -0.03$\pm0.04$ & 0.24$\pm$0.05 & -0.15$\pm$0.05 & 0.16$\pm$0.05 & 0.22$\pm0.05$ \\ \hline 
    $v \sin{I}$ & 1.5$\pm$0.3 & 2.5$\pm0.3$ & 3.28$\pm0.21$ & 3.2$\pm$0.3 & 0.5$\pm$0.3 & 5.9$\pm$0.3 & 2.01$\pm0.3$ \\ \hline 
    Age & 3.2$^{+2.1}_{-3.1}$ & 1.9$^{+1.3}_{-1.8}$ & 3.05$^{+0.92}_{-0.99}$ & 5.2$\pm$1.9 &  11.0$^{+2.4}_{-1.6}$ & 0.4$^{+0.2}_{-0.4}$ & 4.2$^{+1.8}_{-1.9}$ \\ \hline
    Mag$_{NUV}$ & 18.21$\pm$0.04 & 17.63$\pm$0.03 & 17.02$\pm$0.03 & 18.69$\pm$0.05 & 17.03$\pm$0.02 & 17.53$\pm$0.02 & 18.56$\pm$0.05 \\ \hline
    F$_{NUV}$ & 844$^{+45}_{-42}$ & 1306$^{+55}_{-52}$ & 3142$^{+148}_{-141}$ & 1400$^{+95}_{-94}$ & 4147$^{+195}_{-190}$ & 7165$^{+339}_{-316}$ & 1016$^{+76}_{-67}$ \\ \hline
    log$_{10}$(R'$_{hk}$) & -4.476$\pm$0.091 & -4.507$\pm$0.119 & -4.844$\pm$0.146 & -4.781$\pm$0.028 & -4.674$\pm$0.089 & -4.765$\pm$0.056 & -4.455$\pm$0.154 \\ \hline
    P$_{rot_1}$ & 26.37$^{+7.17}_{-4.89}$ & 17.16$^{+3.06}_{-2.62}$ & 16.75$^{+1.41}_{-4.79}$ & 16.62$^{+2.1}_{-5.12}$  & 85.98$^{+132.58}_{-45.19}$ &  9.17$^{+0.65}_{-2.55}$  & 26.73$^{+5.34}_{-9.16}$  \\ \hline
    P$_{rot_2}$ & 29.69$^{+5.00}_{-5.78}$ & 16.20$^{+3.06}_{-0.49}$ & 19.63$^{+3.51}_{-5.39}$ & 25.69$^{+3.15}_{-8.56}$  & 134.42$^{+46.32}_{-58.26}$ & 13.51$^{+5.77}_{-7.40}$ & 27.07$^{+3.96}_{-6.85}$ \\ \hline
    P$_{rot_3}$ & 9.51$\pm$2.74 & 11.65$\pm$4.17 & 25.25$\pm$11.28 & 24.3$\pm$5.12 & 22.8$\pm$6.91 & 24.08$\pm$5.85 & 8.29$\pm$3.8 \\ \hline
    P$_{rot}$ & 28.28$^{+3.99}_{-3.97}$ & 16.93$^{+2.02}_{-1.55}$ & 17.45$^{+2.4}_{-2.85}$ & 20.16$^{+2.88}_{-4.22}$ & 120.1$^{+60.46}_{-37.54}$ & 10.65$^{+3.27}_{-3.01}$ & 25.7$^{+4.16}_{-4.71}$ \\ \ChangeRT{1.6pt}
    M$_p$ & 0.467$^{+0.024}_{-0.023}$& 0.564$^{+0.026}_{-0.025}$ & 0.586$^{+0.033}_{-0.032}$ & 0.47$^{+0.034}_{-0.036}$ & 0.487$^{+0.052}_{-0.054}$ & 0.577$^{+0.057}_{-0.056}$ & 0.65$^{+0.06}_{-0.062}$ \\ \hline
    R$_p$ & 1.119$^{+0.017}_{-0.018}$& 1.232$^{+0.014}_{-0.017}$ & 1.314$^{+0.021}_{-0.025}$ & 1.23$^{+0.027}_{-0.025}$ & 1.177$^{+0.028}_{-0.024}$ & 1.337$^{+0.028}_{-0.033}$ & 1.201$^{+0.033}_{-0.028}$ \\ \hline
    T$_{eq}$ & 1167$\pm$96& 1217$\pm$101 & 1342$^{+111}_{-110}$ & 1350$\pm$112 & 1158$\pm$95 & 1481$\pm$123 & 1230$\pm$102 \\ \hline
    G$_p$ & 9.65$^{+0.59}_{-0.56}$& 9.62$^{+0.52}_{-0.48}$ & 8.8$^{+0.59}_{-0.56}$ & 8.04$^{+0.67}_{-0.71}$ & 9.1$^{+1.0}_{-1.1}$ & 8.36$^{+0.92}_{-0.88}$ & 11.7$^{+1.2}_{-1.3}$ \\ \hline
    $\rho_{p}$ & 0.332$^{+0.023}_{-0.022}$& 0.301$^{+0.019}_{-0.017}$ & 0.258$^{+0.02}_{-0.019}$ & 0.252$^{+0.024}_{-0.025}$ & 0.299$^{+0.037}_{-0.039}$ & 0.241$^{+0.03}_{-0.028}$ & 0.375$^{+0.044}_{-0.047}$ \\ \hline
    I$_p$ & 288$^{+47}_{-45}$& 341$\pm$68 & 504$^{+96}_{-93}$ & 517$^{+127}_{-128}$ & 279$\pm$51 & 748$^{+178}_{-174}$ & 356$^{+93}_{-87}$ \\ \hline
    P$_{orb}$ & 3.3610026 & 3.7648337 & 4.465631 & 3.4252567 & 3.7784022 & 3.3726511 & 4.6058827 \\ 
     & $^{+6.1e-7 -6.3e-7}$ & $^{\pm1.2e-6}$ & $^{\pm1.3e-6}$ & $^{\pm 1.2e-6}$ & $^{\pm 1.6e-6}$ & $^{+2.7e-6 -2.9e-6}$ & $^{\pm 1.1e-6}$ \\ \hline
a/R$_*$ & 11.21$^{+0.13}_{-0.14}$ & 11.33$\pm$0.14 & 10.66$^{+0.14}_{-0.15}$ & 9.13$\pm$0.17 & 11.2$^{+0.17}_{-0.18}$ & 9.22$\pm$0.13 & 11.36$^{+0.2}_{-0.26}$ \\ \hline
t$_0$ & 2454596.43260 & 2455274.99649 & 2455737.93919 & 2456258.06272 & 2456502.72415 & 2457028.58329 & 2457031.95666 \\ 
& $^{+0.000762 -0.00075}$ & $^{+0.00100 -0.00103}$ & $^{+0.00097 -0.00095}$ & $^{+0.00084 -0.00088}$ & $^{\pm0.00102}$ & $^{+0.00103 -0.00101}$ & $^{+0.00025 -0.00026}$ \\ 
\hline
b & 0.195$^{+0.077}_{-0.114}$ & 0.357$^{+0.035}_{-0.042}$ & 0.235$^{+0.067}_{-0.105}$ & 0.724$^{+0.019}_{-0.02}$ & 0.319$^{+0.059}_{-0.072}$ & 0.619$^{+0.027}_{-0.033}$ & 0.395$^{+0.065}_{-0.055}$ \\ \hline
i & 89.0$^{+0.59}_{-0.41}$ & 88.19$^{+0.23}_{-0.2}$ & 88.74$^{+0.57}_{-0.38}$ & 85.45$\pm$0.2 & 88.37$^{+0.38}_{-0.33}$ & 86.15$^{+0.24}_{-0.21}$ & 88.01$^{+0.3}_{-0.38}$ \\ \hline
K & 69.6$^{+2.6}_{-2.7}$ & 75.3$^{+2.5}_{-2.6}$ & 68.7$\pm$3.2 & 62.1$^{+3.8}_{-3.9}$ & 73.1$^{+7.6}_{-7.7}$ & 70.8$\pm$6.7 & 78.3$^{+6.3}_{-6.6}$ \\ \hline
    \enddata
\tablecomments{The top 15 parameters, delineated by a thicker line, are for the host stars. In order, they are mass [M$_{\odot}$], radius [R$_{\odot}$], effective temperature [K], log$_{10}$ of surface gravity [cgs], density [$\rho_{\odot}$], log$_{10}$ of iron to hydrogen abundance relative to the sun [dex], radial velocity [km/s], age [Gy], NUV magnitude, NUV flux from the star at the semi-major axis of the planet [Watts/m$^2$], log$_{10}$ of the Calcium H and K indices [dex], rotational period derive from $v \sin{I}$ (P$_{rot_1}$), photometry (P$_{rot_2}$), R'$_{hk}$ (P$_{rot_3}$), and the adopted rotational period (a weighted mean of the $v \sin{I}$ and photometry periods, P$_{rot}$) [days]. The following 12 parameters are for the corresponding planet and are mass [M$_j$], radius [R$_j$], equilibrium temperature [K], surface gravity [m/s$^2$], density [$\rho_j$], insolation [I$_{\oplus}$], and orbital period [days], semi-major axes relative to the host stars' radii, reference mid transit time transit [BJD], impact parameters, orbital inclinations [degrees], and RV semi-amplitude [m~s$^{-1}$], respectively.}
\end{deluxetable*}

\subsection{Stellar Parameters}\label{sec:StelPara}
We re-derived the stellar parameters of each host star using archival HARPS \citep{Mayor2003} and FEROS \citep{Stahl1999} spectra\footnote{ \href{http://archive.eso.org/eso/eso_archive_main.html}{archive.eso.org}}. Each spectra was homogeneously reduced using \texttt{CERES} \citep{Brahm2017a}, and the stellar parameters ($T_{eff}$, log $g$, [Fe/H], $v \sin{I}$) were derived using \texttt{ZASPE} \citep{Brahm2017b}, as described in detail in \cite{brahm:2019,brahm:2020}. Once \texttt{ZASPE} had obtained the stellar atmospheric parameters, the remaining physical parameters were computed comparing synthetic values generated using PARSEC isochrones \citep{bressan:2012}, and Gaia DR2 \citep{gaia:dr2} parallaxes. For this step we fixed the stellar metallicity to the values obtained by \texttt{ZASPE}, and used $T_{eff}$ from \texttt{ZASPE}  as a prior to obtain posterior distributions for the stellar age, mass, and interstellar extinction using \texttt{emcee} \citep{emcee:2013}. From these values we also derived stellar radii and ${\rm \log g}$, which was then fed back into \texttt{ZASPE} as a fixed parameter. The process described above was iterated until reaching convergence in ${\rm \log g}$. For WASP-55, which has both HARPS and FEROS observations, we weighted averaged the values from each instrument. 

\subsubsection{$R'_{HK}$ Activity Index}\label{sec:R_hk}
We derived $R'_{HK}$ values for each host star from the Ca II $H\&K$ lines in their
HARPS and FEROS spectra, following the methods in \cite{Noyes1984}, and calibrated them to the standard Mount Wilson scale following \cite{Lovis:2011}. To calibrate the $R'_{HK}$ indexes from HARPS and FEROS to the Mount Wilson scale we used six of the seven (all but HD 219834) reference stars used by \cite{Lovis:2011} that have both HARPS and FEROS spectra. For HARPS we found a conversion of the form $S_{MW}=1.118\,\cdot\,S_{HARPS} + 0.0135$, with a 0.0035 fit dispersion. For FEROS we found the conversion $S_{MW}=1.2121\,\cdot\,S_{FEROS} + 0.0072$, with a 0.0275 fit dispersion. 

\subsubsection{Rotational Period}\label{sec:rotP}
We estimated the rotation period, $P_{rot}$, of each star using three proxies: 1)  $v \sin{I}$ from section \ref{sec:StelPara} using eq.
\begin{align}
P_{rot} = 2 \pi R_{*} / v \sin{(i-\lambda)},
\label{equ:prot}
\end{align}
\noindent where R$_{*}$ is the stellar radius, $i$ is the inclination of the planet's orbital axis, and $\lambda$ is the rotation axis alignments --  $\lambda$ has been measured for  WASP-6 and WASP-25 \citep{Gillon2009, Tregloan_Reed2015,Brown2012}, and we estimated it for the remaining stars using probability distributions (see Appendix~\ref{app:lambda}),
2) TESS \citep{Ricker:2014} and ASAS-SN \citep{Shappee:2014,Kochanek:2017} light curves (see Appendix~\ref{app:periodLCs}), and 3) the $R'_{HK}$ indexes derived above, combined with Table 3 and eq. 6 of \cite{Suarez2016}.

The values of $P_{rot}$ obtained via each method are summarized in Table \ref{tab:Sim7Params}. The values obtained with the first two methods are consistent with each other, while the values obtained using $R'_{HK}$ are not fully consistent. Because the periods obtained from the first two methods are more direct measurements, we used the weighted mean from these two methods as our adopted values, reported in Table \ref{tab:Sim7Params}.

\subsubsection{Near-UV Flux}
We derived Near-Ultraviolet (NUV) fluxes for each host star using their Galaxy Evolution Explorer \citep[GALEX; ][]{Bianchi1999} observations. We obtained GALEX NUV magnitudes from \href{https://vizier.cds.unistra.fr/viz-bin/VizieR}{VizieR}, \edit2{which can also be accessed via the Mikulski Archive for Space Telescopes (MAST)\dataset[10.17909/T9H59D]{https://doi.org/10.17909/T9H59D}. The magnitude was converted to total NUV luminosity using} eq. 6 of \cite{Schneider2018}, and Gaia D3 parallaxes, assuming a central mean wavelength of 2267\AA. We then converted those magnitudes to flux density as
\begin{align}
F_{NUV} [\frac{\text{erg}}{\text{s~cm}^2~\text{\AA}}] = 2.06 * 10^{-16} * 10^{\frac{20.08 - m_{NUV}}{2.5}},
\label{equ:Mag2Flux}
\end{align}

\noindent where m$_{NUV}$ is the observed GALEX AB NUV magnitude. This equation was derived from \href{https://asd.gsfc.nasa.gov/archive/galex/FAQ/counts_background.html}{gsfc.nasa.gov}.

\subsection{Planetary parameters} \label{sec:PlantPars}
We re-derived the parameters of each planet using the SPOC\footnote{Science Processing Operations Center \citep{Jenkins2016}} TESS light curves \edit2{downloaded from MAST\dataset[10.17909/fwdt-2x66]{https://doi.org/10.17909/fwdt-2x66} and 
described in Appendix~\ref{app:periodLCs}}, and radial velocity observations from CORALIE \citep{Queloz2000}, HARPS \citep{Mayor2003}, and CYCLOPS \citep{Horton2012}. 
We also used HAT-South \citep{Bakos2013} data with TESS to fit the transit of HATS-29b because of contamination from a background \textit{RR Lyrae} star in the TESS data. After removing that contamination (see Appendix \ref{app:RRLyrae}), the TESS and HAT-South transit light curves were modeled using the same procedure as the other six targets. A table summarizing the RV and photometric data is provided in Appendix \ref{app:periodLCs}.

We jointly fitted the light curves and RVs of each target with \texttt{Juliet} \citep{juliet}, initially assuming circular orbits and quadratic limb darkening coefficients with uniform priors from 0 to 1. All other orbital parameters (P, t$_0$, a/R$*$, R$p$/R$*$, b, i, K, and $\rho_{p}$) were fitted with Gaussian priors set off of the discovery papers' mean and uncertainty values. Then we phase folded the photometric light curves based on the period of that fit, and removed points that were 3$\sigma$ deviant from a moving average of 20 points. This resulted in a few percent of data-points being removed per sector. The clipped data was then used for the final Juliet joint RV-transit fit. 

\begin{figure*}[htb]
    \centering
    \includegraphics[width=1\textwidth] 
    {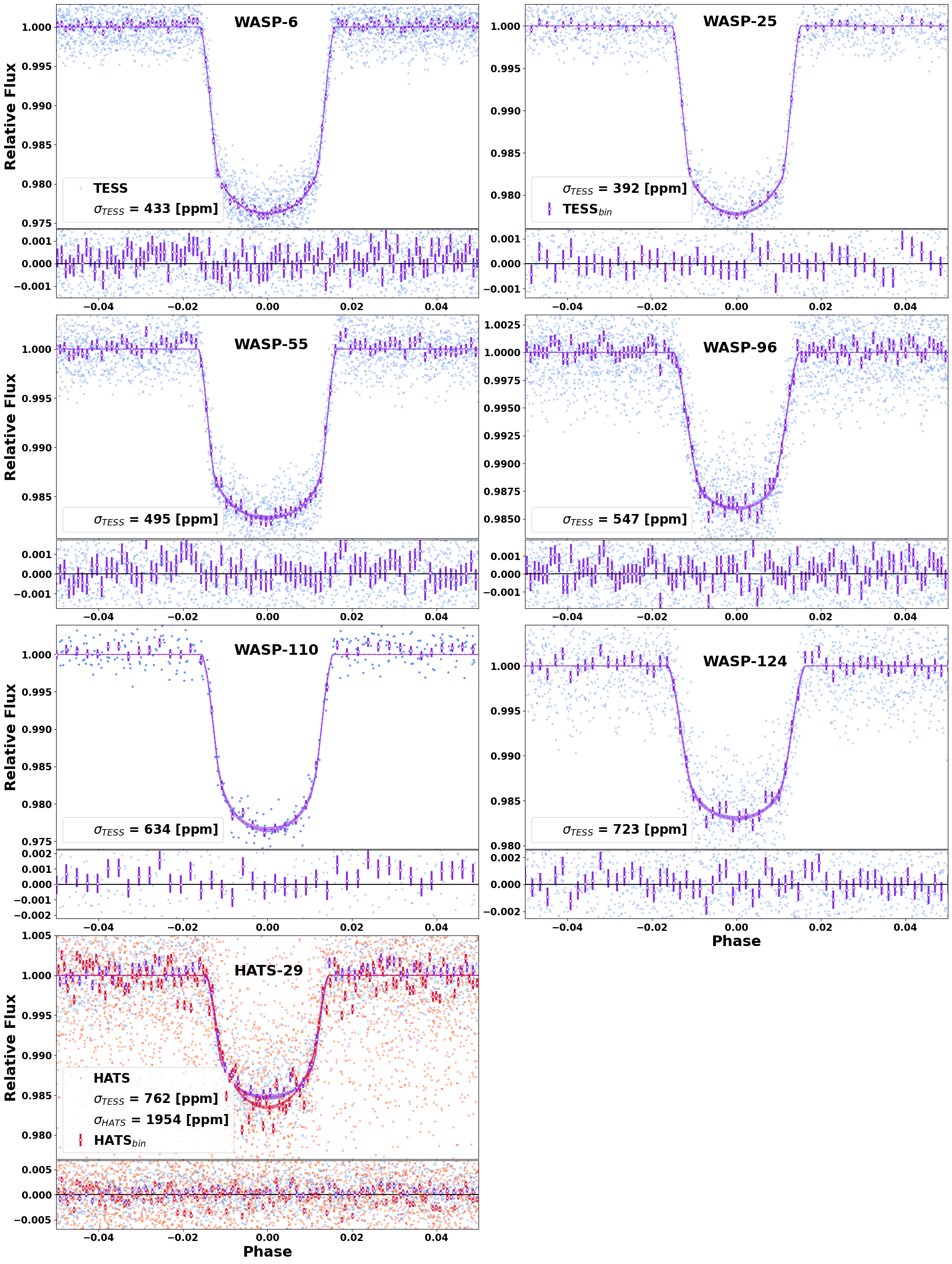}
    \caption{The phase folded transit data for each of the similar seven systems and their  \texttt{Juliet} best fit transit models (dotted lines). 1-$\sigma$ uncertainties of the fits are shaded in the same color as the transit models. The plotted 'bin' data is with 30 points binning, aside for WASP-110, which had 8 point binning.}
    \label{fig:Transits} 
\end{figure*}

\begin{figure*}[htb]
    \centering
    \includegraphics[width=1\textwidth]{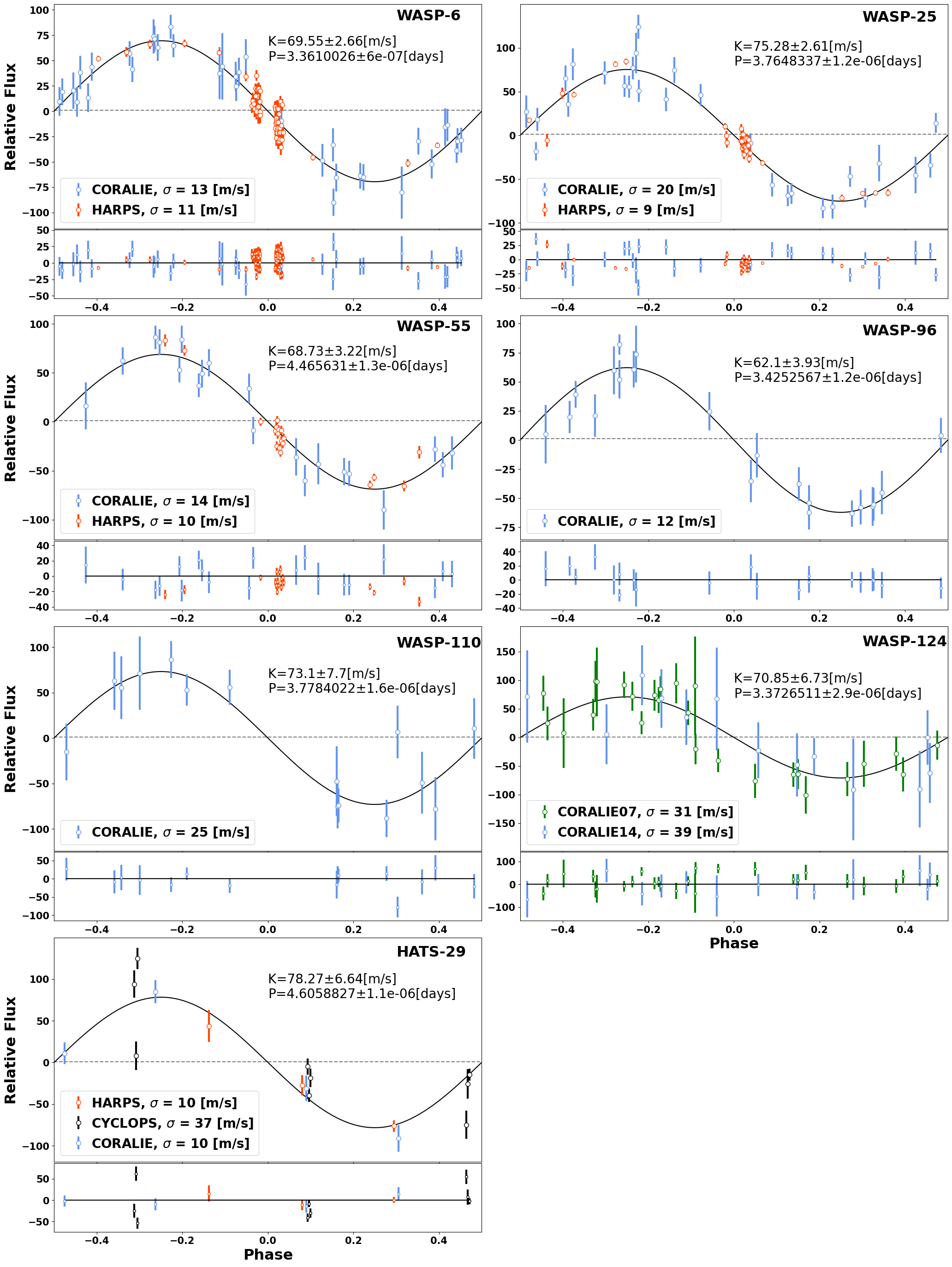}
    \caption{The phase folded RV data of each of the similar seven systems and their \texttt{Juliet} best fit RV models (solid black lines). The HARPS RV measurements that were during transit were omitted. For WASP-124b, 'CORALIE07' and 'CORALIE14' represents observations taken before/after the CORALIE upgrade \citep[see][]{Maxted2016}.}
    \label{fig:RVfits} 
\end{figure*}

The equilibrium temperature of each planet, $T_{eq}$, was computed using eq. 1 from \cite{LopezMorales_2007} and assuming each planet had the same atmospheric albedo and energy redistribution factors of $A_B$ = 0.2$\pm$0.1 and $f$ = 1/3$\pm$0.1\footnote{$A_B$ is based on geometric albedos measured for other hot jupiters \citep[e.g.][]{Mallonn_2019, Adams_2022, Blazek2022}, and $f$ is based on the expectation that gas giants with lower incident flux tend to have more efficient heat redistribution \citep{PerezBecker_2013, Komacek2016, Komacek2017}}. To approximate the insolation levels, ${\rm I_p}$, reaching each planet we calculated the bolometric luminosity assuming the stars emit as blackbodies. To compensate for this oversimplification, we increased the obtained uncertainties in each ${\rm I_p}$  by a factor of three. Given that we can also obtain information about stellar densities from the transits, we weighted averaged the density obtained from the spectral analysis discussed in Section \ref{sec:StelPara} and the density obtained from \texttt{Juliet} to produce the values in Table \ref{tab:Sim7Params}. Our updated parameters are on average 2.7 times more precise than previous literature discovery parameters. Figures \ref{fig:Transits} and \ref{fig:RVfits} show the resulting light curve and radial velocity fits for each planet.

\section{Homogeneous Analysis of Transmission Spectra}\label{sec:TransSpec}

Our last step in the process of obtaining as homogeneous as possible parameters for all seven systems was to re-derive atmospheric parameters for the three planets in the sample with observed transmission spectra. Those are WASP-6b, observed with VLT/FORS2, HST/STIS G430 and G750, and HST/WFC3 G141 between 0.32 and 1.65 $\mu$m \citep{Nikolov:2015, Carter2020}, WASP-96b observed with VLT/FORS2, Magellan/IMACS, and HST/WFC3 G102 and G141 between 0.4 and 1.64 $\mu$m \citep{Nikolov:2018, Yip:2021, Nikolov2022, McGruder2022}, and WASP-110b observed with VLT/FORS2 between 0.4 and 0.83 $\mu$m \citep{Nikolov2021}. 

A retrieval analysis of the WASP-96b data was recently done by \cite{McGruder2022}, using \texttt{Platon} \citep{Zhang_2019PLATON} and \texttt{Exoretrievals} \citep{Espinoza2019}. For consistency, we did a similar analysis for the transmission spectra of WASP-6b and WASP-110b. That is,  with \texttt{Exoretrievals} we tested models including water, potassium, sodium, stellar activity, or scattering features and the different combinations of each. With \texttt{Platon} we tested models with scatters or stellar activity, where \texttt{Platon} assumes equilibrium chemistry and fits for the C/O ratio and planetary metallicity to extrapolate the abundances of atomic/molecular species. ${\rm ln Z}$ Bayesian evidences were used to favor one model over another. We considered a difference in ${\rm ln Z}$ greater than 2.5 between two models to be moderately significant, and greater than 5 to strongly support the higher ln$Z$ model \citep{2008Trotta, 2013Benneke}. 
Additionally, we used Table 2 and eq. 2 from \cite{Rackham:2019}, to limit the contribution of stellar activity to inhomogeneity covering fractions of 4.1 $\pm$ 4.1 $\%$ when both spots and faculae are present.

The priors of each retrieval for both WASP-6b and WASP110b and the Bayesian evidences relative to a flat (for \texttt{Exoretrievals}) or clear (for \texttt{Platon}) spectrum are shown in Appendix \ref{app:retrievalfits}. The highest evidence models for both targets with both retrievals were ones that included activity, which is consistent with what was found in previous analyses \citep{Nikolov:2015, Carter2020, Nikolov2021}. \texttt{Exoretrievals} also found significant evidence for water and a sodium feature in the WASP-6b spectrum, which was not found for WASP-110b. The features found in the WASP-6 data are muted, indicative of high altitude aerosols. Activity could not explain the muted features, in fact, with the unocculted cooler spots that the retrievals find, the sodium signal would be enhanced. This can be seen in Figure 9 of \cite{Carter2020}. Furthermore, the cloud deck pressures of $\sim$ 0.1 bars suggest WASP-6b has substantially more aerosols than WASP-96b, where its models favor a cloud deck pressure of $\sim$ 20 bars. WASP-110b's spectrum is more extreme than WASP-6b's, where all atomic features are missing and a cloud deck pressure of $\sim$ 0.03 mbars is suggested, albeit not well constrained due to the lack of features. 

\section{Search for Trends}\label{sec:trends}

Using the new set of homogeneously derived parameters described in sections \ref{sec:DirParms} and \ref{sec:TransSpec}, we searched for correlations between system parameters and what we define as \textit{aerosol levels} in the transmission spectrum of the planets.

We quantify \textit{aerosol levels} using as proxy the amplitude of the Na I doublet at 5892.9{\AA} in the transmission spectra of WASP-6b, WASP-96b, and WASP-110b. This was calculated as the sum of transit bins within a narrow range centered on the Na I doublet (5862.9 to 5892.9 {\AA}) minus the sum of bins blue-ward and red-ward of this region, while also being outside of the wings of the sodium feature. For WASP-96b this was 4880 to 5380{\AA} and 6200 to 6700{\AA}, but 5340 to 5820{\AA} and 5960 to 6440{\AA} for WASP-6b and WASP-110b which did not have notable absorption wings. 

\begin{figure*}[htb]
    \centering
    \includegraphics[width=1\textwidth]{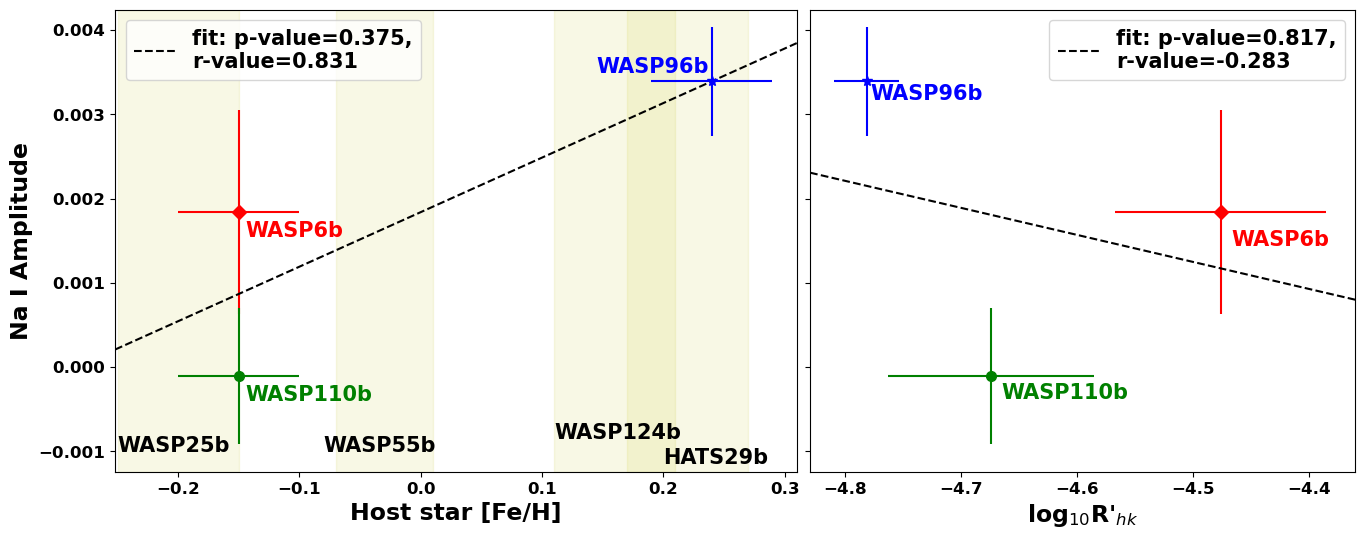}
    \caption{\textbf{Left:} Sodium amplitude versus host star metallicity from the observed WASP-6b (red diamond), WASP-96b (blue star), and WASP-110b (green circle) transmission spectra. The best fit linear regression (black dashes) has a Pearson’s correlation coefficient, \textit{r-value}, of 0.831. The metallicty range WASP-25b, WASP-55b, WASP-124b, and HATS-29b cover are plotted as yellow shaded regions. \textbf{Right:} Same as left, but with Na amplitude versus log$_{10}$(R'$_{HK}$). Here the r-value to the best fit linear regression does not support a correlation of the Na amplitude to log$_{10}$(R'$_{HK}$).}
    \label{fig:CloudCorrelations} 
\end{figure*}

The result of our search for correlations between aerosol levels and system parameters is summarized in Figure \ref{fig:CloudCorrelations}. We find a significant correlation between the amplitude of the Na I feature in the transmission spectrum of the planet and the overall [Fe/H] of the host star. The linear fit to this trend has a Pearson correlation coefficient of r=0.83, corresponding to a 99\% confidence of a found correlation. To examine if there could be a correlation with stellar activity, we used log$_{10}$(R'$_{HK}$) (see Section \ref{sec:R_hk}). The best linear fit to log$_{10}$(R'$_{HK}$) and the Na I signal has r = -0.28, corresponding to about 24\% confidence that there is such a correlation. Therefore, we find no correlation with log$_{10}$(R'$_{HK}$). However, the chromospheric activity measured from log$_{10}$(R'$_{HK}$) is not a direct measurement of total high energy flux \citep[e.g. see][]{Zhang2020,Johnstone2021}, where total high energy emission is likely the more important parameter affecting aerosol formation rates \citep{moses:2011, moses:2013, Fleury:2019}. Thus, we need more direct measurements of the host stars high energy levels (e.g. HST/UVIS or XMM-Newton observations) to confidently rule out such a correlation.

\section{Summary and Conclusions}\label{sec:Sum+Conc}
We have identified seven systems that have very similar characteristics to one another, the '{\it Similar Seven}', where the host star metallicity is the only stark difference between the parameters measured for these systems. Three of the planets in this sample already have transmission spectra observed, and though they have similar parameters, their transmission spectra have widely varying amounts of high altitude aerosols obscuring features. To thoroughly search for correlations between the observed spectra aerosol levels and system parameters, we homogenously reanalyze HARPS and FEROS stellar spectra, and HARPS, CORALIE, and CYCLOPS RV data with TESS and HAT-South transit data to refined the stellar and transit parameters. We found that host star metallicity seems to correlate with the observed aerosol levels, with a 99\% confidence that a linear correlation exists, implying that planets around higher metallicity stars would have lower high altitude aerosol levels. If this holds, it could be explained by the requirement of viable \textit{seed} particles needing to be lofted to high enough altitudes for cloud forming gases to condense on \citep[e.g.][and references therein]{Helling:2008}. The higher metallicity might cause the formed \textit{seed} particles to be more dense and subsequently differentiate lower in the atmosphere. However, given that the potential trend was found with only the three observed transmission spectra, the correlation is tentative, pending on further observations of the other sample planets. 

We also use log$_{10}$(R'$_{HK}$) as a proxy to explore if high energy irradiation could be correlated to the differences in the transmission spectra, given that there are no direct measurements of the stars' high energy emissions. We found no clear signs of a correlation with this parameter. However, correlation to the host star's high energy levels may still be present and require more direct measurements, i.e. with XMM-Newton and/or HST/UVIS.

Regardless of if metallicity or high energy irradiation is found to be a contributing factor to high altitude aerosols in these system, the \textit{similar seven} planets are ideal targets for understanding the unique physical and chemical processes undergoing in these class of planets. This is because the similarity of most parameters act as a controlled sample. This approach of specifically selecting very similar targets should be a common practice in exoplanet atmosphere studies, and has the potential to isolate key physical or chemical phenomena. 

\appendix
\section{Stellar Rotation axis distribution}\label{app:lambda}

\begin{figure*}[htb]
    \centering
    \includegraphics[width=.7\textwidth]{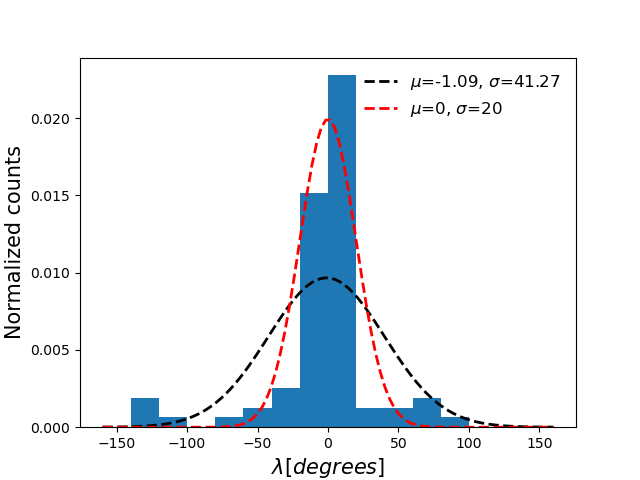}
    \caption{The distribution of observed $\lambda$ values of 80 exoplanet systems with the host star between 5300 and 6300K, obtained through \texttt{TEPCat}\footnote{https://www.astro.keele.ac.uk/jkt/tepcat/obliquity.html}. Here $\lambda$ = 0$^{\circ}$ means the axes are fully aligned. The bulk of the $\lambda$ values are smaller than $\pm 20^{\circ}$ suggesting  alignment of the orbital and spin axes. Two gaussians are overplotted: one with the wide distribution ($\sigma$ = 41.3, black) and one with a narrow distribution ($\sigma$ = 20, red). We adopt a mean distribution of $\lambda = 0 \pm 30^{\circ}$.}
    \label{fig:LambdaDistrib} 
\end{figure*}

For WASP-6 there are two measurements of the rotation axis alignment, $\lambda$, i.e. the angle between the planet's orbital axis and the rotation axis of the star: $\lambda$ = 11$^{+14}_{-18}$~$^{\circ}$, obtained using the Rossiter-McLaughlin effect \citep{Gillon2009} and  $\lambda$ =7.2$\pm$3.7$^{\circ}$ (which we adopt), obtained using occulted star spots \citep{Tregloan_Reed2015}. For WASP-25 \cite{Brown2012} measured a $\lambda$ of 14.6$\pm$6.7$^{\circ}$ via Rossiter-McLaughlin. The other five stars in our sample do not have direct $\lambda$ measurements, so instead we calculated their most likely $\lambda$ values using the distribution of $\lambda$ values measured for G-type stars, i.e. with $T_{eff}$ $\in$ [5300,6300] K, as shown in Figure  \ref{fig:LambdaDistrib}. 75\% of the $\lambda$ values are less that $\pm 20^{\circ}$, suggesting that the bulk of exoplanet systems with G-type host stars are aligned. This is in agreement with the findings of \cite{Triaud2018} (see their Fig. 6). Based on the distribution of values in Fig. \ref{fig:LambdaDistrib}, we adopt a $\lambda = 0 \pm 30^{\circ}$ for the remaning five systems in our sample. The estimated rotation periods for all our targets, computed using eq. \ref{equ:prot}, are listed in Table \ref{tab:Sim7Params}.

\section{Stellar Rotation periods from TESS and ASAS-SN light curves}\label{app:periodLCs}

We downloaded the TESS SPOC \footnote{Science Processing Operations Center \citep{Jenkins2016}} light curves for each target from \edit2{MAST\dataset[10.17909/fwdt-2x66]{https://doi.org/10.17909/fwdt-2x66}}. The TESS sectors and observed number of transits for each target are summarized in Table \ref{tab:jointFitSources}. We also downloaded ASAS-SN \citep{Shappee:2014,Kochanek:2017} time series observations for each target in V- and g-bands, which are treated as separate photometric monitoring campaigns. The number of ASAS-SN observations per photometric band, for each target are also summarized in Table \ref{tab:jointFitSources}.   

We used the time series observations above to estimate the rotation period of each star following a similar analysis to the one described in \cite{McGruder:2020}. For the TESS data, we masked all the in-transit points using the known ephemerides of each planet, to search for photometric modulations of the stars themselves. Before searching for photometric modulations, we binned the data for each target in 3.33-hour bins (16.66-hour bins for WASP-110). Given that the expected rotation periods for all the stars are longer than 10 days, those binning levels should not affect results.  For the ASAS-SN data, we sigma-clipped observations that deviated by more than 3$\sigma$ from the overall mean of each light curve, weighted averaged all observations for a given night (typically 3 observations per night), and removed observations with uncertainties 3 times larger than the mean uncertainty.

To search for the $P_{rot}$ of each star, we jointly modelled all the photometric datasets for each target using \texttt{Juliet}'s \citep{juliet} gaussian processes (GP) semi-periodic kernel, setting the GP characteristic time-scale and period as common terms between all datasets. The other parameters (jitter term, GP amplitude, and GP constant scale term) were specific to the individual datasets. Also, when data from more than one TESS sector were available, we combined them and modeled them together. We modeled activity as a semi-periodic GP instead of using periodograms because it has been found that peaks non-consistent with the rotational period of stars can appear in the latter \citep{Haywood2014,Nava2020}. We assume the main driving factor for the GP period is stellar inhomogeneities coming in and out of view as the star rotates, as such we call this the rotation periods, $P_{rot}$. The priors for $P_{rot}$ were set using the $v \sin{I}$ of each star from section \ref{sec:StelPara}. For all the targets but WASP-124 and HATS-29 we used normal priors with mean and standard deviations near the values derived from $v \sin{I}$. For WASP-124 we used a normal prior truncated at 5 days to prevent sampling of unrealistic low periods driven by the TESS data. For HATS-29 we used a uniform distribution between 17 and 33 days for the same reason.

\edit2{The advantage of the TESS data is the continuous monitoring with high photometric precision. However, given that it only monitors a sector for about 24 days, the long baseline of the ASAS-SN data complements TESS well. This emphasizes the advantage of a joint fit with all photometric data, but we also run the TESS and ASAS-SN data separately (with the same priors) to outline the contribution from each monitoring source. Table \ref{tab:TESSnASASSN_GPfits} shows the period, amplitude, and median absolute deviations (MAD) of both photometric monitoring sources. }

Finally, to ensure that the found rotation periods were not driven by the sampling of the data we tested their window functions, where we used all timestamps of monitoring data but set the flux and uncertainties to 0. Doing this suggested no periodic signal due to the observing cadence. The $P_{rot}$ values for each system are listed in Table \ref{tab:Sim7Params}. 

\begin{deluxetable*}{|c|cc||ccc|ccc|}[htb]
    \caption{Summary of the photometric and RV data} 
    \label{tab:jointFitSources}
 \tablehead{& \multicolumn{2}{|c|}{\bfseries {\large Transit data}} &\multicolumn{3}{|c|}{\bfseries {\large ASAS-SN}} & \multicolumn{3}{|c|}{\bfseries {\large RV data}}}
    \startdata 
 \underline{Star} & \underline{Sectors} & \underline{Transits} & \underline{Filter} & \underline{date range} & \underline{Obs.}  & \underline{Spectrograph} & \underline{Obs.} & \underline{Source} \\ \hline
  \bf{WASP-6}   & 2, 29 & 13  & V & 2013-11-25 to 2018-11-26 & 883 & CORALIE & 35 & \cite{Gillon2009} \\ 
  & & & g & 2017-09-16 to 2022-04-18 & 2073 & HARPS & 55 & \cite{Trifonov2020} \\ \hline
  \bf{WASP-25}  & 10 & 6 & V & 2012-01-24 to 2018-08-20 & 820 & CORALIE & 28 & \cite{Brown2012} \\
  & & & g & 2017-12-21 to 2022-04-18 & 2401 & HARPS & 31 & \cite{Trifonov2020} \\ \hline
  \bf{WASP-55} & 10, 37 & 8 & V & 2012-02-17 to 2018-08-19 & 961 &  CORALIE & 20 & \cite{Hellier2012}   \\ 
  & & & g & 2017-12-16 to 2022-04-19 & 2156 & HARPS & 19 & \cite{Trifonov2020} \\ \hline
  \bf{WASP-96} & 2, 29 & 13 & V & 2014-04-30 to 2018-09-24 & 921 & CORALIE & 21 & \cite{Hellier:2014}$^*$ \\ 
  & & & g & 2017-09-05 to 2022-02-10 & 2592 & & & \\ \hline
  \bf{WASP-110} & 27 & 6 & V & 2014-04-29 to 2018-09-24 & 916 &  CORALIE & 15 &  \cite{Anderson2014}$^*$  \\ 
    & & & g & 2017-09-05 to 2022-04-18 & 2286 & & & \\ \hline
  \bf{WASP-124}  & 1 & 8 & V & 2014-04-30 to 2018-09-19 & 1577 & CORALIE & 39 &  \cite{Maxted2016}$^*$  \\ 
    & & & g & 2017-09-07 to 2022-04-18 & 4217 & & & \\ \hline
  \bf{HATS-29} & 13 & 6 & V & 2014-05-17 to 2018-09-24 & 989 & HARPS & 3 & \cite{Espinoza2016}\\
  (\it{HAT-South} & & & g & 2017-10-03 to 2022-04-18 & 2165 & CYCLOPS & 9 & \cite{Espinoza2016}\\
  \it{data:})& -- & 23 & & & &CORALIE & 4 & \cite{Espinoza2016}\\
    \enddata
 \tablecomments{The HAT-South photometric data for HATS-29 (first columns, last row) was acquired from \cite{Bakos2013}. \cite{Trifonov2020} reanalyzed archival HARPS data using \href{https://www2.mpia-hd.mpg.de/homes/trifonov/HARPS_RVBank.html}{\texttt{SERVAL}} \citep{Zechmeister2018}. The number of observations is denoted 'Obs.', and is the unbinned/unclipped observations for the ASAS-SN data.}
\tablenotetext{*}{data obtained through \href{https://dace.unige.ch/dashboard/}{DACE}}
\end{deluxetable*}

\begin{deluxetable*}{|c|ccc|cccc|cc|}[htb]
    \caption{Summary of TESS and ASAS-SN photometric monitoring \texttt{Juliet} fits} 
    \label{tab:TESSnASASSN_GPfits}
    \tablehead{{\bfseries {\large Star}} & \multicolumn{3}{|c|}{\bfseries {\large TESS}} &\multicolumn{4}{|c|}{\bfseries {\large ASAS-SN}} &\multicolumn{2}{|c|}{\bfseries {\large Joint}}}
    \startdata 
             & \underline{MAD}  & \underline{Period} & \underline{Amplitude} &  \underline{MAD$_g$} & \underline{MAD$_v$} & \underline{Period} & \underline{Amplitude} & \underline{Period} & \underline{Amplitude}\\
    & [ppm]  & [days] & [ppm] & [ppm]  & [ppm]  & [days] & [ppm] & [days] & [ppm] \\ \hline
   WASP-6    & 348.5  & 28.93$^{+6.63}_{-6.9}$  & 0.23 & 2407.3  & 2266.2  & 28.37$^{+5.58}_{-7.19}$ & 1383.1 & 29.69$^{+5.00}_{-5.78}$ & 57.3 \\
   WASP-25   & 376.8  & 18.11$^{+2.82}_{-2.99}$  & 187.9 & 1847.6  & 1459.3 & 16.11$^{+1.73}_{-0.63}$ & 947.2 & 16.20$^{+3.06}_{-0.49}$ & 199.5   \\
   WASP-55   & 241.2  & 19.46$^{+3.77}_{-3.87}$  & 132.4  & 1516.3  & 2494.4 & 15.54$^{+6.36}_{-3.21}$ & 1181.2 & 19.63$^{+3.51}_{-5.39}$  & 49.1 \\
   WASP-96   & 230.0  & 19.85$^{+5.05}_{-5.84}$  & 58.7   & 1955.89   & 1761.3  & 15.38$^{+10.78}_{-2.78}$  & 703.8  & 25.69$^{+3.15}_{-8.56}$ & 98.1    \\
   WASP-110  & 330.6  & 112.2$^{+52.4}_{-47.6}$    & 48.0 & 2063.9   & 2220.8  & 132.9$^{+43.6}_{-71.1}$ & 488.1 & 134.42$^{+46.32}_{-58.26}$  & 58.8     \\
   WASP-124  & 447.3  & 12.04$^{+4.84}_{-6.0}$  & 256.4   & 1427.5   & 1441.3  & 10.98$^{+4.44}_{-3.21}$ & 714.6  & 13.51$^{+5.77}_{-7.40}$  & 340.1    \\
   HATS-29   & 737.5  & 26.06$^{+4.79}_{-5.78}$  & 78.3   & 1775.7   & 2340.9  & 20.78$^{+9.063}_{-0.585}$  & 1654.5   & 27.07$^{+3.96}_{-6.85}$ & 90.1     \\
    \enddata
 \tablecomments{The MAD is for the binned data. The amplitdues are obtained by using \texttt{scipy.optimize.minimize} \citep{Virtanen2020} to fit a sine curve to the data phase folded on the \texttt{Juliet} best fit period. The subscripts g and v on the ASAS-SN MAD correspond to the g and V band filters.}
\end{deluxetable*}

\section{HATS-29 TESS Light Curve Decontamination}\label{app:RRLyrae}
\begin{figure*}[htb]
    \centering
    \includegraphics[width=1\textwidth]{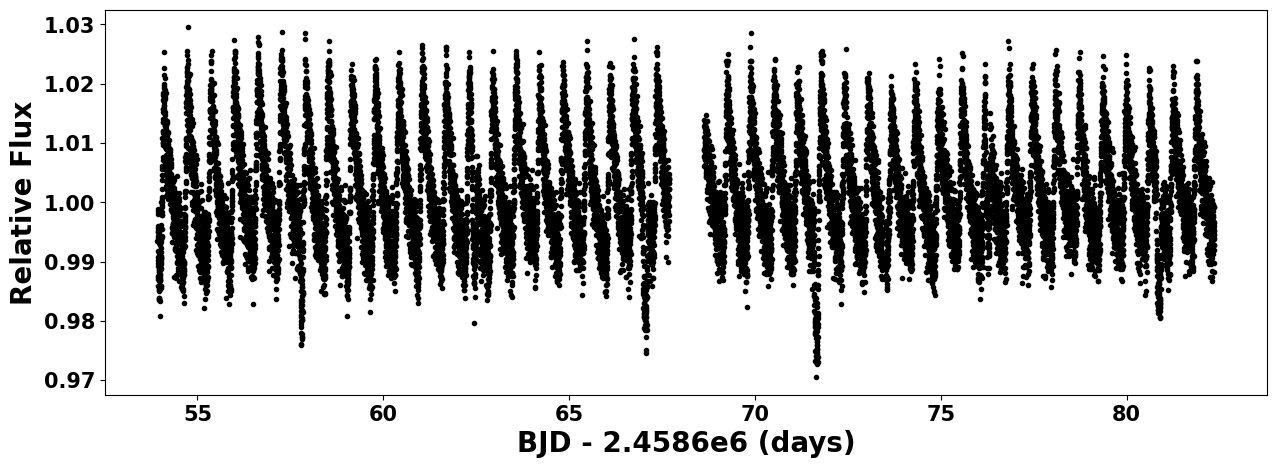}
    \includegraphics[width=1\textwidth]{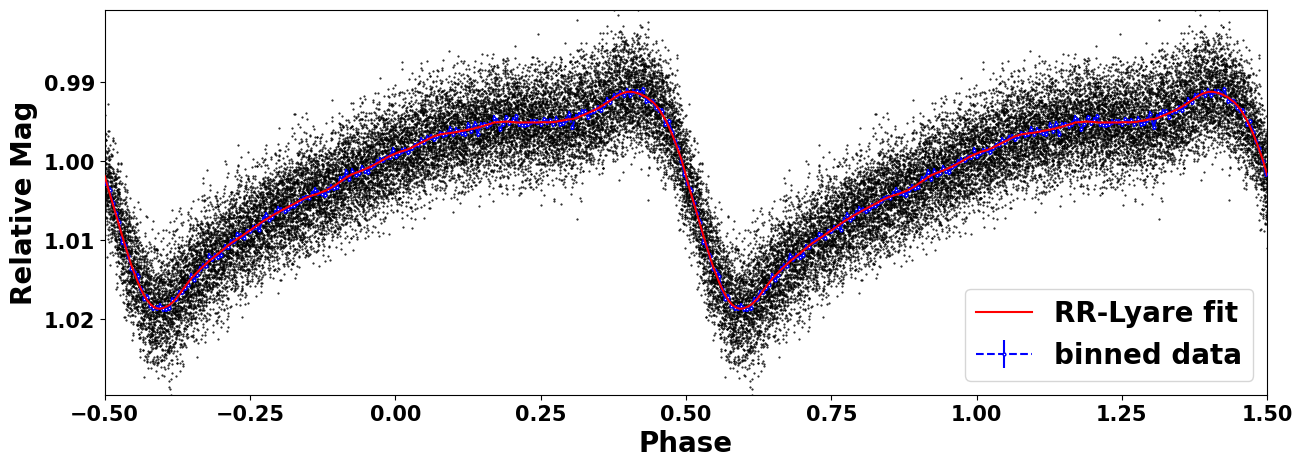}
    \includegraphics[width=1\textwidth]{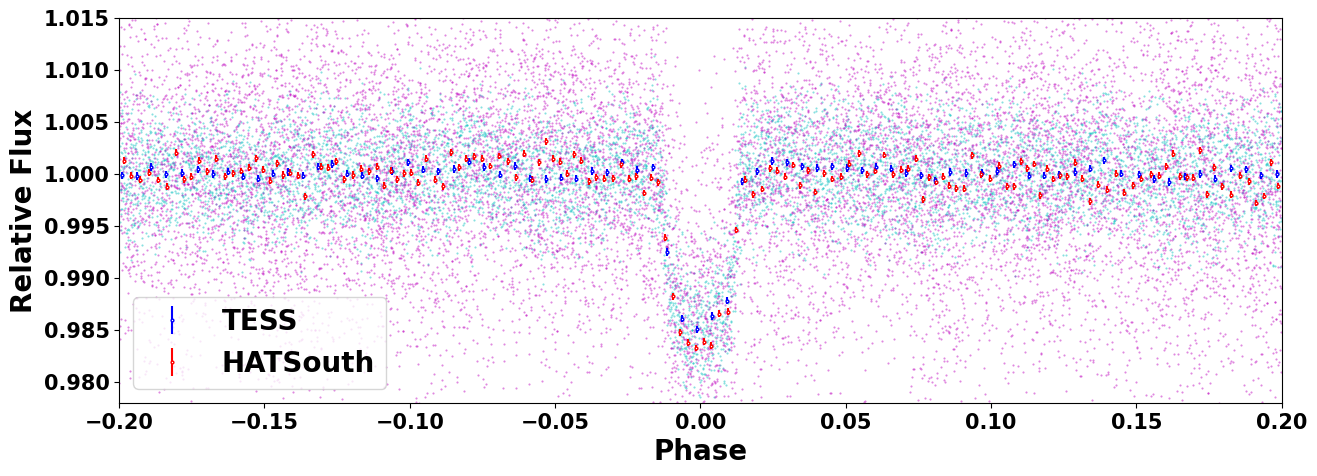}
    \caption{\textbf{Top:} The TESS SPOC of TIC 201604954, extracted from \edit2{MAST\dataset[10.17909/fwdt-2x66]{https://doi.org/10.17909/fwdt-2x66}}, which was observed in sector 13. From visual inspection one can see that the RR-Lyrae oscillations dominate, with 43 complete oscillations observed. \textbf{Middle:} The same data after removing the HATS-29b transits and phase folding on a period of 0.631343 days, which corresponds to the background RR-Lyrae's period. Here the black are all phase folded observations, the blue is the data binned by 100, and the red is the \href{https://scipy.github.io/old-wiki/pages/Cookbook/SavitzkyGolay}{\texttt{savitzky\_golay}} smoothed fit. \textbf{Bottom:} The TESS data after modeling out the RR-Lyrae features, doing the same sigma clipping done for every other target (see Section \ref{sec:PlantPars}). Here cyan is the unbinned data and the blue is the same data binned by 100. Overplotted is the HAT-South data, where the magenta and red are the unbinned and binned data, respectively. For both light curves we used a P and t$_0$ of 4.60588 and 2457031.957, respectively. }
    \label{fig:HATS29dat} 
\end{figure*}

Because of the low image resolution of TESS, the light curve of HATS-29 is contaminated by a background \textit{RR Lyrae} star (see top panel of Figure \ref{fig:HATS29dat}). We isolated the RR-Lyrae signal by first excluding the in transit data (using transit parameters from \citep{Espinoza2016}), then applying a Lomb-Scargle periodigram \citep{Lomb1976, Scargle1982} analysis to this out-of-transit data to find a period of 0.631 days. Next, we phase folded the data to the 0.631~d period, binned the data by 100 points, and finally smoothed the binned data with \texttt{scipy.signal.savgol\_filter} \citep{Virtanen2020} \footnote{We used an older version, '\href{https://scipy.github.io/old-wiki/pages/Cookbook/SavitzkyGolay}{\texttt{savitzky\_golay}},' which is the same algorithms before it was included in scipy.} that had a window set to 51 and order of 10. Our phase folded data and corresponding RR-Lyrae model can be seen in the second panel of Figure \ref{fig:HATS29dat}. We then subtracted (in magnitude space) the best RR-Lyrae light curve model from the TESS data, and reduced the corrected TESS data with the same procedures of the other TESS observations (see Appendix \ref{app:periodLCs}). 

HAT-South \citep{Bakos2013} has public light curves for HATS-29, which we downloaded from the survey's website \footnote{hatsouth.org}, and is not contaminated by the background star. We compared the best fit \texttt{Juliet} transit with TESS against a fit with the HAT-South data. Upon confirming that the transit parameters - aside from transit depth, which one would expect to differ due to the different photometric bands - were consistent with each other, we ran a joint \texttt{Juliet} fit with all the transit and RV data to obtain our final planetary parameters of this system. See the bottom panel of Figure \ref{fig:HATS29dat} for an overlay of the TESS and HAT-South data.

\section{Atmospheric Retrieval Results}\label{app:retrievalfits}
Table \ref{tab:Priors} has the priors used for each model and the $\Delta \ln Z$ of each retrieval run are in Table \ref{tab:W6_W110_LnZs}

\begin{deluxetable*}{|C|C|C|C|C|C|}[htb]
    \caption{The priors for \texttt{Exoretrievals} and \texttt{PLATON}}
    \label{tab:Priors}
    \tablehead{\multicolumn{3}{|c|}{\bfseries {\large Exoretrievals}} &\multicolumn{3}{|c|}{\bfseries {\large PLATON}}}
    \startdata 
    \textbf{parameter}                          & \textbf{function} & \textbf{bounds} & \textbf{parameter}           & \textbf{function} & \textbf{bounds} \\ \hline
    \text{reference pressure (P\textsubscript{0}, bars)}                            & \text{log-uniform}      & \text{-8 to 3}        &  \text{reference pressure (P\textsubscript{clouds}, Pa)}           & \text{log-uniform} & \text{-3.99 to 7.99} \\ \hline
    \text{planetary atmospheric}                & \text{uniform}          & \text{600 to 1800K}   & \text{planetary atmospheric} & \text{uniform} & \text{600 to 1800K} \\
    \text{temperature (T\textsubscript{p})}                      &                         &                    & \text{temperature (T\textsubscript{p})}        &                & \\ \hline
    \text{stellar temperature}& \text{uniform} & \text{ T\textsubscript{eff}-240 to T\textsubscript{eff}+240K}& \text{stellar temperature }& \text{gaussian}         & $\mu$\text{=T\textsubscript{eff}}, $\sigma$\text{=150K} \\ 
    \text{(T\textsubscript{occ})}   &      &                & \text{(T\textsubscript{star})}   &  & \\ \hline
    \text{stellar heterogeneities}              & \text{uniform}          & \text{T\textsubscript{eff}-3000 to T\textsubscript{eff}+3000K}& \text{stellar heterogeneities} & \text{uniform} & \text{T\textsubscript{eff}-3000 to T\textsubscript{eff}+3000K} \\
    \text{temperature (T\textsubscript{het})}   &                         &  & \text{temperature (T\textsubscript{spot})} &    &   \\\hline
\text{heterogeneity covering}                        & \text{gaussian}          & $\mu$\text{=0.041, }$\sigma$\text{=0.041}          & \text{heterogeneity covering}                                      & \text{gaussian} &$\mu$\text{=0.041, }$\sigma$\text{=0.041}  \\ 
    \text{fraction (f\textsubscript{het})}      &                         &               & \text{fraction (f\textsubscript{spot}) }                                      &                   &                   \\ \hline
    \text{haze amplitude ($a$)}                 & \text{log-uniform}         & \text{-30 to 30}           & \text{scattering factor}                                           & \text{log-uniform} & \text{-10 to 10}\\ \hline
    \text{haze power law (}$\gamma$\text{)} 
    & \text{uniform }         & \text{-14 to 4}            & \text{scattering slope (}$\alpha$\text{)} 
    & \text{uniform} & \text{-4 to 14}\\ \hline
    \text{log cloud absorbing} & \text{uniform} & \text{-80 to 80}        & \text{metallicity (Z/Z}$_{\odot}$)     & \text{log-uniform} & \text{-1 to 3}\\
    \text{cross-section (}$\sigma$\text{\textsubscript{cloud})} &                   &      &                                            &  & \\ \hline
    \text{trace molecules'}       & \text{log-uniform}      & \text{-30 to 0 }             & \text{C/O}                                                         & \text{uniform} & \text{0.05 to 2}\\
    \text{mixing ratios}       &     &            &                                                        &  & \\ \hline
    \text{reference radius factor} ($f$)   & \text{uniform }         & \text{0.8 to 1.2}            & \text{1\,bar, reference radius (R\textsubscript{0})}                       & \text{uniform} & \text{ R\textsubscript{p}-.2R\textsubscript{p} to
    R\textsubscript{p}+.2R\textsubscript{p}}\\ \hline
    \enddata
\tablenotetext{}{\textbf{Note.} These priors were set to allow for a wide parameter space to be surveyed, but contained within physical regimes. Not all parameters were included in each model fit (see Tab.~\ref{tab:W6_W110_LnZs}). We used 5000 live points for all runs. For further description of the parameters of \texttt{Exoretrievals}, please refer to the Appendix D of \citet{Espinoza2019}. T\textsubscript{eff} is the effective temperature of the host star, which is 5438K and 5392K for WASP-6 and WASP-110, respectively. $\gamma$ is the exponent of the scattering slope power law, where $-4$ is a Rayleigh scattering slope. $\alpha$ is the wavelength dependence of scattering, with 4 being Rayleigh. $f$ is a factor multiplied by the inputted planetary radius to produce the reference radius, i.e. R$_0=f$R$_p$, R$_p$ is the radius of the planet, corresponding to 1.119R$_j$ and 1.177R$_j$ for WASP-6b and WASP-110b, respectively.}
\end{deluxetable*}

\begin{deluxetable*}{|l|C|C|C|C|C|C|C|C|C}[h!]
    \caption{$\Delta$ln~Z for \texttt{Exoretrievals} (left) and \texttt{PLATON} (right) retrievals}
    \label{tab:W6_W110_LnZs}
    \tablehead{\multicolumn{7}{|c|}{\bfseries {\large Exoretrievals}} &&\multicolumn{2}{|c|}{\bfseries {\large PLATON}}}
    \startdata 
      \textbf{Model:}  &  \text{flat}  &  $H_2O$  &  $Na$  &  $K$  &  $H_2O +Na$ &  $H_2O + K +Na$  && \textbf{Model:} &\\ \hline
       \textbf{WASP-6b:} \\
      clear &	0.0 &	1.47 &	3.79 &	70.32 &	70.26 &	71.7 && \text{clear} & 0.0\\
      scatterers &	--- &	69.96 &	70.14 &	69.95 &	69.15 &	69.9  && \text{scattering} & 10.09\\
      activity &	84.21 &	93.85 &	88.03 &	90.53 &	96.07 &	97.44 && \text{activity} & 13.49\\
      Both &	--- &	81.1 &	82.35 &	81.59 &	81.49 &	81.97 && \text{Both} & 12.91 \\ \hline
      \textbf{WASP-110b:} \\
      clear &	0.0 &	0.15 &	-1.0 &	-0.97 &	-0.14 &	-0.57 && \text{clear} & 0.0\\
      haze+clouds &	--- &	-3.21 &	-3.09 &	-3.12 &	-3.46 &	-3.6  && \text{scattering} & 2.93\\
      activity &	4.35 &	2.69 &	2.33 &	2.39 &	2.2 &	1.73 && \text{activity} & 4.08\\
      Both &	--- &	-0.08 &	-0.15 &	-0.32 &	-0.43 &	-1.0 && \text{Both} & 4.15 \\
    \enddata 
    \tablenotetext{}{\textbf{Note.} The $\Delta$ln~Z values are relative to a clear (and flat for \texttt{Exoretrievals}'s case) spectrum with the WASP-6b (\textbf{top}) spectrum that included the VLT/FORS2, HST/STIS, and HST/WFC3 data, and the WASP-110b (\textbf{bottom}) spectrum consisting of the VLT/FORS2 data. For WASP-6b the retrievals with water and sodium were heavily supported by \texttt{Exoretrievals}, where including potassium did not make a significant difference in $\Delta$ln~Z. The \texttt{PLATON} models that included scattering and activity were equally supported as the models with just activity. For WASP-110b the models with activity were supported, with \texttt{Exoretrievals} finding no significant contribution from atomic/molecular species.}
\end{deluxetable*}

\begin{acknowledgments}
We thank the anonymous referee for helpful comments to the manuscript.
This work has been supported by the National Aeronautics and Space Administration's Exoplanet Research Program via grant No. 20-XRP20\_2.0091.
We thank N. Espinoza for providing access to \texttt{Exoretrievals}, S. Blanco-Cuaresma for continuous support using \texttt{iSpec}, E. Shkolnik for helpful discussion regarding GALEX data, and V. DiTomasso for discussion regarding analysis of RV data. 
We also appreciate the support from the NSF Graduate Research Fellowship (GRFP), grant No. DGE1745303.
RB and AJ acknowledge support from ANID -- Millennium  Science  Initiative -- ICN12\_009. AJ acknowledges additional support from FONDECYT project 1210718. RB acknowledges support from FONDECYT project 11200751.

\end{acknowledgments}

%

\vspace{5mm}
\facilities{Magellan:Baade (IMACS), Smithsonian Institution High Performance Cluster (SI/HPC), HST(STIS/WFC3), 
All-Sky Automated Survey for Supernovae (ASAS-SN), VLT(FORS2),TESS, ESO La Silla 3.6m (HARPS), Swiss 1.2-metre Leonhard Euler Telescope (CORALIE), Anglo-Australian Telescope (CYCLOPS), MPG/ESO2.2(FEROS), and Gaia spacecraft}


\software{Astropy \citep{astropy2013}, corner \citep{corner2016}, Matplotlib \citep{matplotlib2007}, NumPy \citep{numpy2020}, Multinest \citep{multinest2009}, PyMultiNest \citep{2014BuchnerPyMultiNest}, SciPy \citep{scipy2020}, batman \citep{Kreidberg2015_batman}, george \citep{Mackey2014_george}} dynesty \citep{Speagle_2020Dynesty}, \texttt{Platon} \citep{Zhang_2019PLATON}, Juliet \citep{juliet}, CERES \citep{Brahm2017a}, ZASPE \citep{Brahm2017b}, iSpec \citep{Blanco-Cuaresma2014, BlancoCuaresma2019}




\bibliographystyle{yahapj}
\bibliography{References}



\end{document}